\def\Mv{$M_{\rm v}$}

\def\0BMV{$(B~-~V)_{\rm 0}$}
\def\BMV{$B~-~V$}
\def\Z{$Z$}
\def\Feh{[Fe/H]}

\def\Zsun{$Z_\odot$}

\def\simgt{\lower.5ex\hbox{$\; \buildrel > \over \sim \;$}}
\def\simlt{\lower.5ex\hbox{$\; \buildrel < \over \sim \;$}}

\documentclass{aa}
\usepackage{graphicx}
\begin{document}
   \thesaurus{04(10.19.1;    
	      10.11.1;    
	      10.19.2)}    
\title{Kinematics and age of stellar populations in the solar neighbourhood
from Hipparcos data\thanks{Based on
data collected by the Hipparcos ESA consortium}}
%
%
\author{V. Caloi \inst{1}, D. Cardini \inst{1}, F. D'Antona \inst{2},
M. Badiali \inst{1}, A. Emanuele \inst{1} and I. Mazzitelli \inst{1}}

   \offprints{V. Caloi}
   \mail{caloi@victoria.ias.rm.cnr.it}

   \institute{Istituto di Astrofisica Spaziale C.N.R., Via Fosso del
	      Cavaliere, I-00133 Roma, Italy
	      \and
	      Osservatorio Astronomico di Roma, I-00040 Monte Porzio, Italy
	     }

   \date{Received 23 December 1998; accepted 21 September 1999}

   \titlerunning{Kinematics and age of local populations }
   \authorrunning{V. Caloi et al.}
   \maketitle

   \begin{abstract}
{The large sample of main sequence stars with Hipparcos parallaxes and
estimates of the radial velocity has been used to study the relations among
kinematics, age and heavy element content for the solar neighbourhood. The
samples with $V$ $>$ -30 km $\rm s^{-1}$, although including stars as old as
$10^{10}$ yr, have a very young component (age $\simgt$ $10^7$ yr), and have
a metallicity close to the solar one. Most of the stars with space velocity
component $V$ $<$ -40 km $\rm s^{-1}$ have a minimum age of about 2 $10^9$
yr. Stars with $V$ \simlt\ -80 km $\rm s^{-1}$ share a common age \simgt\
$10^{10}$ yr. The distribution in \Feh\ of these latter objects suggests a
decrease in metal content with increasing $\vert V \vert$ till \Feh\ $<$ -2.
For $V$ $<$ -180  km $\rm s^{-1}$ only the metal poor component
([Fe/H] $<$ -0.7) is found.  The common age for large space velocity
suggests that no substantial age spread exists in the inner halo, at least
in the local sample.}
    \end{abstract}


\section{Introduction}

The recognition of the existence of stellar populations differing for age,
chemical composition, spatial distribution and kinematical properties has
been the fundamental breakthrough in the knowledge of galactic structure, and
the basis for models of galaxy formation and evolution (see, e.g., the
review by Majewski 1993). The large subsample of the Hipparcos catalogue
for which radial velocity and proper motions are available appears a
promising tool to attempt a description of the local behaviour of
stellar populations.

The derived database is largely incomplete and heterogeneous, and various
steps will be necessary to be able to reach a conclusion. With due
caution, it is possible to form an idea of the variations in age and heavy
element abundance with variations in kinematics, i.e., in the components
$U,V,W$, of star space velocity in the Galaxy. The necessary information on
metal content is obtained (for part of the selected objects) from existing
catalogues. On the basis of these data and of theoretical isochrones we check
that differences in \BMV\ correspond to the expected differences in \Feh:
this will allow to interprete the observed colour of the main sequence
location in terms of metal content.

We examine also the behaviour of stellar orbit eccentricities as function of
space velocity, which gives a picture of the local galactic structure
consistent with the one suggested by star distribution as function of metal
content.

The results we obtain are valid for the solar neighbourhood; in the
limits in which this region is representative of the body of the galaxy, such
results can give hints on galactic structure on a larger scale.

\section{The velocity and metallicity samples}

We extracted from the Hipparcos Catalogue (ESA 1997) the single objects with
an error on the parallax $\sigma_ \pi$/$\pi$ $\leq$ 0.15 and $\sigma(B~-~V)$
$\leq$ 0.1, for which estimates of the radial velocity were found in the
literature (Turon et al. 1992; Barbier-Brossat et al. 1994; Duflot et al.
1995a,b; Fehrenbach et al. 1996, 1997), for a total of 9972 objects. When
more than one value was found for an object, either the most recent or the
one with the smallest error was chosen. We accepted recognized periodic
variables (total number 267). We also considered single objects for
which $\sigma_\pi$/$\pi$ $<$ 1 (total number about 18000, 810 periodic
variables) only for comparison purpose. Known or suspected non-single
stars have been excluded on the basis of Hipparcos Catalogue indications.

\begin{figure*}
{\caption{CM diagrams for Hipparcos stars with
metallicity estimates in Cayrel de Strobel et al. (1997). Two reference
theoretical main sequences with $Z=0.01$\ and $Z=0.02$\ are shown. The shift
to the red of main sequence stars with increasing \Z\ is on the average in
agreement with theoretical expectations. The empty main sequence region for
\Mv\ $\simlt$ 3 mag is due to the peculiar surface metal abundance
phenomenon.}}
\vspace{1cm}
\label{zdep}
\end{figure*}

The velocity sample obtained above is not complete and suffers from many
biases, which will be discussed in the following section.

From the velocity sample we extracted those objects having estimates
of the metal abundance, on the basis of the catalogue by Cayrel de Strobel
et al. (1997), to which we added 124 stars from Carney et al. (1994), for a
total of 1541 stars ($\sigma_\pi$/$\pi$ $\leq$ 0.15). Cayrel's catalogue is a
compilation from several sources, so that its data are rather inhomogeneous.
Furthermore, for many objects more than one value of \Feh\ is given: in this
case, we selected the most recent value, in the hypothesis that it is the
most reliable one. Of course this is not always correct, but it should be at
least statistically true.

In first instance, the sample has been used to determine the mean location of
the main sequence as a function of \Feh, in order to make sure that the \BMV\
of a main sequence star in the velocity sample is generally well related to
its metal content. We used the ($V$,\BMV) magnitudes and colours in the
Johnson $UBV$ photometric system provided by the Hipparcos Catalogue. The
absolute magnitudes have been calculated without any allowance for
interstellar reddening.

This subject has been addressed by many authors on the basis of Hipparcos
data (Bartkevicius et al. 1997, Cayrel et al. 1997, Lebreton et al. 1997,
Badiali et al. 1997). Lebreton et al. examined in particular the question of
the main sequence position as function of the chemical composition. They
found that the separation in colour among main sequences with different metal
content is lower than expected theoretically, a result supported by
Bartkevicius et al. and Badiali et al..

\begin{figure*}
\caption[]{CM diagrams for stars in the survey sample with $V$\ in different
bins. Isochrones are shown as reference; ages from top are 2 and 5 $\times
10^7$, 1, 2 and 5  $\times 10^8$, 1,  2, 5 and 10 $\times10^{9}$ yr
(composition $Y=0.27$, \Z=0.017; see Appendix for references). In the first
three diagrams $\sigma_ \pi$/$\pi$ $\leq$ 0.15,
in the other three $\sigma_ \pi$/$\pi$ $<$ 1.
The diagrams equivalent to the first two figures, but
with $\sigma_ \pi$/$\pi$ $<$ 1 do not show differences in the turnoff
age; the velocity bin with -30 $< V <$ -20 is shown also with $\sigma_
\pi$/$\pi$ $<$ 1 to exemplify their similarity. The less reliable data
(0.15 $<$ $\sigma_ \pi$/$\pi$ $<$ 1) are
identified by the large error bars.
}\label{fig2}
\end{figure*}

On the basis of the present, larger sample we obtain the behaviour in Fig.
1, where objects of different metallicity are shown together with
two main sequence loci with $Z=0.01$\ and 0.02, taken as reference lines.
Contrary to what found in precedence, on the average main sequence loci move
progressively toward redder colours with increasing \Z. We attribute this
difference to the larger size of the present sample and to a better estimate
of the metal content, based on the latest edition of Cayrel et al.'s
catalogue (1997).

A more quantitative approach is described in the Appendix, where we compare
the expected value of $\Delta (B-V)/\Delta$ \Feh\ with the observed one.
With some caution, it appears possible to follow changes in populations of
different metal content through changes in main sequence loci. This will
allow to interprete the CM diagrams we shall construct for the stars for
which we have the space velocity but not [Fe/H].

\section{ Age and kinematics }

In first instance we examine the sample given by the Hipparcos Survey,
complete to between the 8th and 9th mag for stars with 0 $<$ \BMV\ $<$ 0.8,
according to galactic position (Perryman et al. 1989); objects with
\BMV\ $>$ 0.8 are complete to a slightly lower magnitude.

The survey limiting magnitude is sufficient to provide a good definition of
the turn-off region for the oldest stars (\Mv\ about 4 $-$ 5 mag) up to
galactic velocities close to those of the halo (see Figs. 2 and 3).

In Table 1 we give the distribution of main sequence stars in the
Hipparcos Survey (about 23400 objects) with \Mv\
and $\sigma_\pi$/$\pi$, considering for the latter quantity the intervals 0
to 0.15, 0.15 to 1 and larger than 1. Thirteen objects have been excluded
because they did not fulfill the condition on the colour error
$\sigma(B~-~V)$ $\leq$ 0.1. As expected, the objects with the largest errors
($\sigma_\pi$/$\pi$ $>$ 1) have \Mv\ $<$ 0: we shall not consider them, since
both their magnitude and velocity would turn out totally unreliable (their
number is less than 150). For \Mv\ $>$ -4 mag the majority of the objects
has $\sigma_\pi$/$\pi$ $<$ 1, for \Mv\ $>$ 1 mag the majority has
$\sigma_\pi$/$\pi$ $\leq$ 0.15.
The turnoff age at \Mv\ = 1 is $\simeq 5 \times 10^8$ yr, thus the
stars with large parallax errors are mostly younger. Consequently, we can
not derive strong conclusions for {\it smaller} ages, but the older ages
will be safely known.

\begin{table}[t]
\caption[ ]{Distribution with \Mv\ of main sequence survey stars, separated
according to the error on the parallax.}
\renewcommand{\arraystretch}{1.4}
\begin{tabular}{cc cc cc cc}
\hline\noalign{\smallskip}
 \Mv   &$\sigma_\pi$/$\pi$ $\leq$ 0.15 & 0.15 $<$ $\sigma_\pi$/$\pi$ $<$ 1
  & $\sigma_\pi$/$\pi$ $\geq$ 1  \\
\noalign{\smallskip}
\hline
\noalign{\smallskip}
-6$\div$-5    & 1 & 6 & 40 \\
-5$\div$-4    & 2 & 30 & 54 \\
-4$\div$-3    & 13 & 105 & 86 \\
-3$\div$-2    & 31 & 326 & 49 \\
-2$\div$-1    & 108 & 849 & 8 \\
-1$\div$0     & 378 & 1378 & 2 \\
0$\div$1      & 1078 & 1918 & 1 \\
1$\div$2      & 2553 & 2116 & 0 \\
2$\div$3      & 3935 & 1403 & 0 \\
3$\div$4      & 4049 & 388 & 0 \\
4$\div$5      & 1905 & 26 & 0 \\
5$\div$6      & 442 & 1 & 0 \\
6$\div$7      & 80 & 0 & 0 \\
7$\div$8      &  5 & 0 & 0 \\
8$\div$9      & 3 & 0 & 0 \\
9$\div$10     & 1 & 0 & 0 \\
10$\div$11    & 1 & 0 & 0 \\
\hline
\end{tabular}
\renewcommand{\arraystretch}{1}
\label{Tab01}
\end{table}

Main sequence objects have been defined as those either with \Mv\ $>$ 4 or
\BMV\ $<$ 0.14, or to the left of the segment between the two points (\BMV\
= 0.14, \Mv\ = 0) and (\BMV\ = 0.9, \Mv\ = 4).

We are interested in a subset of the survey, that is, in the stars with
radial velocity estimate for which it is possible to evaluate the space
velocity components $U,V,W$. We assume for the velocity of the
Sun with respect to the local standard of rest ($V_{\rm LSR}$ = 220
km $\rm s^{-1}$, Kerr \& Lynden-Bell 1986)
the values: $U_\odot$ = 11.0 km $\rm s^{-1}$, $V_\odot$ = 5.3 km $\rm
s^{-1}$, $W_\odot$ = 7.0 km $\rm s^{-1}$ (Binney et al. 1997).
Considering only the main sequence,
we find 7996 stars with $v_{\rm r}$ (out of about 23400). Table 2 is
the equivalent of Table 1 for stars with $v_{\rm r}$. We added the
percentage, for each magnitude bin, of the stars with $v_{\rm r}$ estimates
and $\sigma_\pi$/$\pi$ $<$ 1, with respect to the total number of main
sequence stars in the survey (in the same magnitude bin).

Bright stars (-5 $\simlt$ \Mv\ $\simlt$ 0) with $v_{\rm r}$ are present
in percentage $\simgt$ 40\%, while for 0 $\simlt$ \Mv\ $\simlt$ 4 the
percentage is lower ($\simgt$ 25\%).
For \Mv\ $\simgt$ 6 $-$ 7 mag, we shall make use of the
complete catalogue to improve the statistics.

\begin{table}[t]
\caption[ ]{Distribution with \Mv\ of main sequence survey stars with
estimate of $v_{\rm r}$, separated according to the error on the
parallax. In the last column the percentage of stars with $v_{\rm r}$ and
$\sigma_\pi$/$\pi$ $<$ 1, with respect to the total number of main sequence
stars in that magnitude bin. }
\renewcommand{\arraystretch}{1.4}
\begin{tabular}{cc cc cc cc cc}
\hline\noalign{\smallskip}
 \Mv   &$\sigma_\pi$/$\pi$ $\leq$ 0.15 & 0.15 $<$ $\sigma_\pi$/$\pi$ $<$ 1
  & $\sigma_\pi$/$\pi$ $\geq$ 1 & \%  \\
\noalign{\smallskip}
\hline
\noalign{\smallskip}
-6$\div$-5    & 1 & 6 & 31 & 15 \\
-5$\div$-4    & 2 & 30 & 32 & 37 \\
-4$\div$-3    & 13 & 91 & 34 & 51 \\
-3$\div$-2    & 31 & 262 & 20 & 70 \\
-2$\div$-1    & 108 & 489 & 3 & 62 \\
-1$\div$0     & 344 & 551 & 0 & 51 \\
0$\div$1      & 732 & 471 & 0 & 40 \\
1$\div$2      & 1102 & 321 & 0 & 30 \\
2$\div$3      & 1146 & 157 & 0 & 24 \\
3$\div$4      & 1080 & 19 & 0 & 25 \\
4$\div$5      & 587 & 2 & 0 & 31 \\
5$\div$6      & 210 & 0 & 0 & 47 \\
6$\div$7      & 61 & 0 & 0 & 76 \\
7$\div$8      &  5 & 0 & 0 \\
8$\div$9      & 3 & 0 & 0 \\
9$\div$10     & 1 & 0 & 0 \\
10$\div$11    & 1 & 0 & 0 \\
\hline
\end{tabular}
\renewcommand{\arraystretch}{1}
\label{Tab01}
\end{table}

\subsection{ From one velocity bin to the other. The survey }
We shall consider CM diagrams relative to survey stars both with $\sigma_
\pi$/$\pi$ $\leq$ 0.15 and with $\sigma_\pi$/$\pi$ $<$ 1.

We stress that stars with large errors on the parallax (0.15 $<
\sigma_\pi$/$\pi$ $<$ 1) are used only to check that a sample that includes
almost all the main sequence survey objects with velocity estimates (see
Table 2) gives the same results as the reduced sample with
$\sigma_\pi$/$\pi$ $\leq$ 0.15. Such a check is admittedly an approximate
one, since the errors on \Mv\ and $v_r$\ when $\sigma_\pi$/$\pi$ $>$ 0.25 are
unconfortably large. Still the comparison will allow to show that the sample
with $\sigma_\pi$/$\pi$ $\leq$ 0.15 gives a good description of the whole
survey main sequence population with velocity estimates (except for the
velocity bin -40 $< V <$ -30). This improves the confidence in the results
obtained with the reduced sample.

The main points we have to address are: i) is each diagram representative
of the population with a given velocity?; ii) are the differences among the
diagrams real or do they depend on the sample selection?

For point i), it is not necessary that the sample be complete under some
respect: it is sufficient that it is able to delineate faithfully the
evolutionary loci actually present in the chosen stellar sample, even if not
with the correct population ratio in the various evolutionary phases. Since
we are interested in the {\it minimum} age of a given sample, we have to make
sure that the main sequence is populated at {\it all} the ages occurring at a
given velocity. In other words, we shall have to make sure that the {\it
absence} of main sequence objects at a certain magnitude is real and not the
product of poor statistics.

We shall consider the CM diagrams obtained for star samples with increasing
values of $\vert V \vert$ (Figs. \ref{fig2}, \ref{fig5} and \ref{fig7}),
examining them in relation one to the other.

We start from the first velocity bin (0 $<$ $\vert V \vert$ $<$ 10 km $\rm
s^{-1}$). In this sample the main sequence is well populated from well
below the turn-off of the oldest stars present (about magnitude 4) to a
luminosity corresponding to a minimum age of few $10^7$ yr. Such age may be
overestimated, since our sample can miss some of the youngest stars.

We proceed examining the CM diagrams for decreasing values of the $V$
component. The turn-off region for the oldest stars and the lower main
sequence remain largely similar down to $V$ about -60 $-$ -80, and the
higher main sequence remains well populated up to about few $10^7$ yr as long
as $V$ $>$ -30 km $\rm s^{-1}$.

We interprete the CM diagrams in the velocity bins $V$ $>$ -30 km $\rm
s^{-1}$ as describing a population with the same minimum ($\leq 10^7$ yr) and
maximum age ($\geq$ $10^{10}$ yr). Most of the objects are found at the right
of the \Zsun\ isochrone, suggesting the absence of a substantial number of
metal poor stars.

Exam of the total sample with $\sigma_\pi$/$\pi$ $<$ 1 does not change this
interpretation. In fact
we repeat the CM diagram relative to -30 $< V <$ -20 km $\rm s^{-1}$
considering objects with $\sigma_\pi$/$\pi$ $<$ 1 to show that it actually
carries the same information found in the $\sigma_\pi$/$\pi$ $\leq$ 0.15
case, both for the high and the low regions of the main sequence.

The CM diagram relative to the sample -40 $< V <$ -30 km $\rm s^{-1}$
is the most ambiguous. When the stars with $\sigma_ \pi$/$\pi$ $\leq$ 0.15
only are considered, the CM diagram (not shown) exhibits a strong increase in
the age for a large part of the population, up to about 1 $-$ 2 $10^9$ yr.
\footnote{In this respect, it is important to remember that bright stars are
not discriminated against in a magnitude limited sample such as the survey
is, and that the majority of bright stars (\Mv\ $<$ 0) in the survey has a
radial velocity estimate (Table 2).}
On the contrary, there are apparently many young stars in the sample with
$\sigma_ \pi$/$\pi$ $<$ 1 (Fig. \ref{fig2}), but their large errors
preclude a safe conclusion on the age.

\begin{figure*}
\caption[]{ CM diagrams for stars with the indicated $V$ and different
origin: i) from the survey and $\sigma_\pi$/$\pi$ $<$ 1 ( which
coincides, for main sequence stars with $M_v \geq 3$, with the distribution
from the survey with $\sigma_\pi$/$\pi$ $\leq$ 0.15); ii) from
the complete catalogue and $\sigma_\pi$/$\pi$ $<$ 1, iii) from the complete
catalogue and $\sigma_\pi$/$\pi$ $\leq$ 0.15. The turn off level is the same
in the three diagrams. Isochrones correspond to 5 $10^9$ and $10^{10}$ yr
(composition $Y$ = 0.27, \Z\ = 0.017, see Appendix for references). }
\label{fig5}
\end{figure*}

Moving to the velocity interval -60 $<$ $V$ $<$ -40 km $\rm s^{-1}$, we find
a sharp decrease in the main sequence population for \Mv\ $\simlt$ 3 mag, the
total number of objects being comparable to the one relative to the sample
-40 $< V <$ -30 km $\rm s^{-1}$. Fig. 2 shows the sample having
$\sigma_ \pi$/$\pi$ $<$ 1, in which it can be seen that the very few
stars with large errors are found only in the upper main sequence.

\begin{figure}
\caption{Distribution of the sample randomly extracted from the -60 $< V <$
-40 sample (continuous line) and of the observed -80 $< V <$ -60 sample
(dotted line, see text).}
\label{fig4}
\end{figure}
Considering the
sample with $\sigma_ \pi$/$\pi$ $\leq$ 0.15 (CM not shown) the age increase
is confirmed.
The bulk of the sample with -60 $<$ $V$ $<$ -40 km $\rm s^{-1}$ appears
older than 2 $10^9$ yr.

\begin{figure}
\hspace{0.1cm}
\caption{CM diagram for $\vert V \vert$ $<$ 10 from the complete Catalogue
and $\sigma_\pi$/$\pi$ $\leq$ 0.15. The lower main sequence appears well
populated and redder than the solar metallicity track. }
\label{fig6}
\end{figure}

Increasing the drift velocity to -80 $<$ $V$ $<$ -60 km $\rm s^{-1}$ (Fig.
\ref{fig5}, left panel), we find the main sequence populated only at
magnitudes \Mv\ $\simgt$ 4, with the lower main sequence almost empty. The
estimated minimum age is of at least 5 $10^9$ yr.

Because of the substantial difference in total star number between the
latter two CM diagrams, we performed
a check on the reality of the change in the age structure between
these two velocity samples, by randomly reducing the
number of main sequence members in one diagram (-60 $-$ -40), to the
number in the other one (-80 $-$ -60) and comparing the main sequence
distributions in $M_v$ (samples with $\sigma_ \pi$/$\pi$ $\leq$ .15).
The result is shown in Fig. \ref{fig4}: the random sample confirms the
intrinsic difference in the turn off location between the two velocity
samples.

The -80 $< V <$ -60 km $\rm s^{-1}$ sample is very scantily populated for
\Mv\ $>$ 5 mag: the survey is no more sufficient to describe the lower main
sequence.

\subsection{ The complete Hipparcos sample }

We need to take into account the complete Hipparcos catalogue, which reaches
fainter magnitudes than the survey. Objects at these high magnitudes have a
heterogeneous origin. The targets with \Mv\ $>$ 4 $-$ 5 have been selected
mainly from catalogues like Gliese (1969, 1974), LHS
(Luyten 1976) and NLTT (Luyten 1979); from the Michigan Spectral Survey (Houk
\& Cowley 1975, Houk 1978, Houk 1982, Houk \& Smith-Moore 1988); for their
high galactic latitude, and on the basis of spectral peculiarities and
subdwarf characteristics (Perryman et al. 1989). So the choice was
based largely on high proper motion and peculiarities like weak metal lines,
but also very nearby stars (Gliese catalogue 1969, 1974 and Gliese and
Jahreiss extension 1979), stars at an estimated distance smaller than 40 pc,
and red dwarf stars have been included for \Mv\ $>$ 5 mag.

\begin{figure*}
\caption[]{ CM diagrams for stars with the indicated $V$ from the complete
catalogue with $\sigma_\pi$/$\pi$ $\leq$ 0.15. Isochrones of $10^{10}$ yr
correspond to \Z\ = 0.017, 4 $10^{-3}$ and 6 $10^{-4}$ ($Y$ = 0.27 and
0.23, respectively).}
\label{fig7}
\end{figure*}

How are the properties of the lower main sequence population affected
by the observational selection? Proper motion catalogues favour objects with
high proper motions. Notwithstanding this preference, the Hipparcos catalogue
is such that, even for $\vert V \vert$ $<$ 10 km $\rm s^{-1}$ (see Fig.
\ref{fig6}), the main sequence locus is well populated up to \Mv\ about 10
mag by stars of solar composition. If normal stars are present in a number
sufficient to describe the lower main sequence for low $\vert V \vert$, their
absence - or presence - at larger $\vert V \vert$ can be considered
meaningful.

A possible bias is given by the fact that a sample chosen on kinematical
basis would tend to miss metal poor stars without halo kinematical
characteristics, such as the low metallicity, low eccentricity stars observed
by Norris et al. (1985) and by Norris (1986). Such an occurrence is at least
partially avoided by the presence of targets chosen on the basis of spectral
peculiarities, among which are weak-lined objects (supposedly metal weak).
These spectroscopically selected objects can exhibit in principle any space
velocity, and could populate the subdwarf sequence at any velocity bin.

We show (Fig. \ref{fig5}) three CM diagrams relative to -80 $< V <$ -60
km $\rm s^{-1}$: one based on objects from the survey with
$\sigma_\pi$/$\pi$
$<$ 1, and two from the complete catalogue, both with $\sigma_\pi$/$\pi$ $<$
1 and $\leq$ 0.15. They give equivalent information on the minimum age of the
velocity bin population, but, as mentioned before, the survey sample has
very
few stars below the turn off, so it is not appropriate to investigate the
lower main sequence. We shall refer to the diagram relative to the complete
catalogue and $\sigma_\pi$/$\pi$ $\leq$ 0.15, which contains the same
information as the one with $\sigma_\pi$/$\pi$ $<$ 1 in a more clear form;
the few bright stars with large errors on the parallax confirm the turn off
at about the 4th mag. A minimum age of at least 5 $10^9$ yr appears
appropriate for this velocity bin.

\begin{figure*}
\caption[]{As in Fig. \ref{fig2}, for positive values of the $V$
component.}\label{fig8} \end{figure*}

For larger velocities, we chose the intervals in $V$ which gave the least
dispersion in colour, to follow as closely as possible the evolution of the
average main sequence colour with kinematics. The magnitude level of the turn
off does not change appreciably from $V$ $\simeq$ -80 to $V$ $\simeq$ -500
km $\rm s^{-1}$, while the average main sequence colour does (Fig.
\ref{fig7}).
For -110 $< V <$ -80 km $\rm s^{-1}$ most of the stars are
located on the red side of the 4 $10^{-3}$ track, but a few {\it bona fide}
subdwarfs are present close to the 6 $10^{-4}$ track. For -180
$< V <$ -110 km $\rm s^{-1}$ one finds objects more or less uniformly
distributed from the solar track to the 6 $10^{-4}$ one, while for $V <$ -180
km $\rm s^{-1}$ all the objects are found on the blue side of the 4
$10^{-3}$ track.

\begin{figure}
\caption[]{ Distributions in \Feh\ of stars with decreasing rotational
velocity.}
\label{fig9}
\end{figure}

For completeness the CM diagrams for positive $V$ values are shown in Fig.
\ref{fig8}; the complete catalogue and $\sigma_\pi$/$\pi$ $\leq$ 0.15 have
been considered.

Similar diagrams can be obtained for the $U,W$ components (see later).

\section{ Age, metal content and kinematics in the solar neighbourhood }
\subsection{ Age and metal content }

The behaviour of minimum age with space velocity of the solar
neighbourhood sample contained in the Hipparcos catalogue can be put in
relation to current ideas on galactic structure and kinematics.

For -30 $<$ $V$ $<$ 10 km $\rm s^{-1}$ the CM diagrams suggest a common
minimum
age of about $10^7$ yr. The stars involved belong likely to the local thin
disk population; we have not the sensitivity to observe the Parenago
discontinuity (e.g. Binney et al. 1997), which is in any case not large
enough to affect the discussion.

When $V$ $<$ -30 km $\rm s^{-1}$, at first we notice the disappearance of
large part of the young component of the thin disk population,
and the
appearance of a population with a minimum age of about 1.5 $-$ 2 $10^9$
yr. Its rotational velocity is similar to the one of the thick disk (lagging
velocity with respect to the local standard of rest of 30 $-$ 80 km $\rm
s^{-1}$, Gilmore \& Wyse 1985). At a rotational velocity of about 140 km
$\rm
s^{-1}$ ($V$ = -80 km $\rm s^{-1}$), only objects older than 5 $10^9$ yr are
found, with no substantial subdwarf presence. The minimum metal content,
judged in relation to the solar main sequence track, is of \Z\ about 4
$10^{-3}$.

For $V$ $<$ -80 $-$ -100 km $\rm s^{-1}$ we can interprete the observed
population as one of almost constant age ($\geq$ $10^{10}$ yr), with an
increasing importance of the subdwarf component with decreasing rotational
velocity. For $V$ $<$ -180 km $\rm s^{-1}$ only objects bluer than the \Z\ =
4
$10^{-3}$ track are present; it is not possible to examine more in detail the
relation between metallicity and the $V$ component of these subdwarfs through
the \BMV\ colour, which is not sensitive enough to differences in \Z.

\begin{figure*}
\hspace{0.8cm}
{\caption[]{Distributions - normalized to the star number in the plot - of
four metallicity samples ([Fe/H] as indicated $\pm$ 0.15) as function of
$U,V,W$. Full line: complete sample; dotted line: survey sample.}
\label{fig10}
\vspace{2cm}}
\end{figure*}

We exploited the metallicity sample to investigate the relation between $V$
component and metal content. For each of the velocity bins in Fig.
\ref{fig2}, \ref{fig5} and \ref{fig7}, we obtained the distribution of CM
diagram members with \Feh; they are shown in Fig. \ref{fig9} (we considered
all stars, not only main sequence objects, to improve the statistics). In the
first, low velocity bins the percentage of stars with a metal estimate is of
about 12 \% (with respect to the objects with that velocity and
$\sigma_\pi$/$\pi$ $\leq$ 0.15), and it increases steadily reaching 76 \% in
the -520 $< V <$ -180 bin.

The local metal poor ($Z$ $<$ 4 $10^{-3}$) component apparently
becomes important for $V <$ -110 km $\rm s^{-1}$, but still in presence of a
substantial more metal rich population up to the solar metallicity; it
becomes dominant for $V <$ -180 km $\rm s^{-1}$. The change in
the average main sequence \Z\ between the last two CM diagrams cannot be
ascribed to a kinematic bias, since all objects with $V$ $<$ -110 belong to
the high velocity galactic component.

A similar analysis can be performed with respect to the space velocity
components $U,W$. The results are resumed in Table 3, where characteristic
minimum ages are associated to velocity intervals.

\subsection{Metal content and kinematics}

In Fig. \ref{fig10} we report the distribution of the metallicity sample with
the space velocity components for four values of \Feh\ (+0.3, 0, -0.3 and
-0.6 $\pm$ 0.15). The distribution of survey stars from the same sample is
also shown (dotted line). The distributions are normalized to the total star
number in each plot. The comparison between the complete sample and the
survey one suggests that no strong bias is introduced by the selective
addition of lower main sequence objects to the magnitude complete survey
sample. The metallicity sample is not complete, but its size (more than 1500
stars) makes the information from it worth discussing.

In the case of the +0.3 sample, 90\% of the sample stars has values
of $\vert U \vert$ $\leq$ 40 km $\rm s^{-1}$; for the sample with \Feh\ =
-0.6, such a percentage is 38\%; for \Feh\ = 0 and -0.3, we have 80\%
and 63\%, respectively. For the $V$ component, values comprised between +10
and -30 km $\rm s^{-1}$ amount to 76\%, 67\%, 57\% and 31\%, from +0.3 to
-0.6, respectively.

These features suggest that the samples with \Feh\ = +0.3, 0.0 and -0.3 share
a common kinematic distribution, peaked at $U$ and $V$ about zero; in the
case of the $V$ component, all three samples show a tail towards negative
values.

Stars with \Feh\ = -0.6 apparently belong to a population with a less marked
rotation on the galactic plane. Perhaps this reflects a composition
difference between the disk (thin and thick) component and the (local) inner
halo. The distribution of the vertical component $W$ supports the previous
findings: the three more metal rich samples exhibit distributions peaked at
$W$ = 0, while the \Feh\ = -0.6 sample shows a mild concentration about
zero. In percentages: considering $\vert W \vert$ $<$ 20 km $\rm s^{-1}$ we
have 91\%, 87\%, 68\% and 48\% with \Feh\ from +0.3 to -0.6, respectively.

\begin{table}
\caption[ ]{Approximate minimum turn--off ages for characteristic velocity
groups.}
\renewcommand{\arraystretch}{1.4}
\begin{tabular}{cl@{}|cl@{}|cl@{}}
\hline\noalign{\smallskip}
 $\vert U \vert$& $\tau_{min}$&
 $V$  & $\tau_{min}$    &$\vert W \vert$& $\tau_{min}$    \\
\noalign{\smallskip}
\hline
\noalign{\smallskip}
 0  $-$ 40 & $\leq$ 2 $10^7$  &
 -30  $-$ 10 & $\leq$ 2 $10^7$    &   0  $-$ 15 & $\leq$ 2 $10^7$  \\
 $>$ 60  &  $\geq$ 2 $10^9$   &
 $<$ -40 &  $\geq$ 2 $10^9$  &  $>$ 30 &  $\geq$ $10^9$ \\
 $>$100 &  $\geq$ 10 $10^9$ &
 $<$ -80 & $\geq$ 10 $10^9$ &    $>$ 60&  $\geq$ 10 $10^9$ \\
\hline
\end{tabular}
\renewcommand{\arraystretch}{1}
\label{Tab01}
\end{table}

The tail towards negative $V$ velocities can be intepreted in the usual
scheme as due to the thick disk, presenting a lag or drift from the
flattened, highly rotational velocity thin disk (Str\"omberg 1925, Gilmore
\& Reid 1983, Gilmore \& Wyse 1985, Wyse \& Gilmore 1986).

\begin{figure*}
{\caption{Distribution of the eccentricity versus the
$W$ component for four metal contents (the indicated ones $\pm$ 0.15).}}
\vspace{1cm}
\label{eccen}
\end{figure*}

\subsection{Metal content and eccentricity}

We obtained an estimate of the eccentricity of our velocity and metallicity
sample adopting the Eggen, Sandage and Lynden-Bell (ELS) galactic potential
(Eggen et al. 1962), which considers the $U$ and $V$ components only and
therefore computes the eccentricity of the orbit projected onto the galactic
plane, assuming the independence of the motion on this plane from the
perpendicular one. This approximation of the galactic potential has been
used by many authors beside Eggen et al. (e.g., Carney et al 1990, Chiba
\& Yoshii 1998). It has been demonstrated that the use of this potential
overestimates the eccentricity (Yoshii \& Saio 1979).

We adapted the ELS potential to the updated values of the distance of the
Sun from the galactic center $R_\odot$ and of the circular velocity
$V_{\rm LSR}$ at the Sun position, using the values of 8.5 kpc and
220 km $\rm s^{-1}$, respectively (Kerr \& Lynden-Bell 1986).
We checked that a change in these parameters of the order of their errors
($\pm$\ 10\%) does not influence substantially the results: the variations
rarely reach $\pm$\ 0.04 and the square root of their quadratic mean is $<$
0.03.

In Fig. 10 the behaviour of the eccentricity is shown as function
of $W$, for the four values of \Feh\ considered above. When \Feh\ = 0, 76\%
of the objects have eccentricities $\leq$ 0.2; for \Feh\ = -0.3 the
percentage becomes 63\%, and for \Feh\ = -0.6, 28\% (for \Feh\ = +0.3, the
percentage is 86 \%). Comparing this behaviour with the one in Fig.
\ref{fig10}, one finds again that objects with \Feh\ = -0.6 ($\pm$ 0.15) have
a motion in the Galaxy largely not confined on the galactic plane. The
eccentricities of the sample with \Feh\ = -0.3 being similar to the ones
relative to \Feh\ = 0, the change at \Feh\ = -0.6 appears rather sudden.

\section{ Discussion and conclusions }

Some kinematical features of the local population have been identified.
The first row in Table 3 gives the values of the space velocity components
for which very young stars are found. The presence of low age objects and
the rapidly rotating configuration may be taken to define the thin disk.

For $V$ $<$ -40, $\vert U \vert$ $>$ 60 and $\vert W \vert$ $>$ 30 km $\rm
s^{-1}$ the minimum stellar age is about 2 $10^9$ yr. The disappearance of
the younger stellar component may be assumed to mark the transition to the
thick disk.

For $V$ $<$ -80, $\vert U \vert$ $>$ 100 and $\vert W \vert$ $>$ 60 km $\rm
s^{-1}$ the stellar population exhibit a common age $\geq$ $10^{10}$ yr.
For increasing values of $\vert U,V,W \vert$ no substantial increase in age
is observed, while the average location of the main sequence shifts more
and more to the blue. The metal poor component (\Z\ $<$ 4 $10^{-3}$) is best
isolated by the condition $V$ $<$ -180 km $\rm s^{-1}$ (halo population).

For what concerns the chemical composition, we notice the departure from a
rapidly rotating configuration (-30 $< V <$ 10 km $\rm s^{-1}$) at \Feh\
about -0.6; this kinematical change is possibly related to the age increase
mentioned before, which takes place at similar values of this space
velocity component.

The common age for large space velocity suggests that no substantial age
spread exists in the inner halo, at least in the local sample. Hints of a
progressive decrease of the metal content with increasing $\vert U,V,W \vert$
come from the distributions with \Feh\ at various $V$: this point, together
with the absence of age spread at large $U,V,W$, is more in favour of a model
in which the bulk of the inner halo formed in a continuous coherent process
- collapse of a gaseous mass, as advocated by Eggen et al. (1962) and Sandage
\& Fouts (1987) - than a model in which accretion and merging are
prevailing. Undoubtly episodes of merging have taken, and are taking, place
(as exemplified by the case of the dwarf ellitpical galaxy in Sagittarius,
Ibata et al. 1994), but their effects have not been enough as to cancel the
overall kinematical and chemical structure left over by the collapse and
spin-up of the original protogalaxy, at least in the inner halo region.

\acknowledgements{{\sl ``Hipparchus numquam satis laudatus...ausus rem etiam
deo inprobam, adnumerare posteris stellas...organis excogitatis, per quae
singularum loca atque magnitudines signaret, ut facile discerni posset...an
omnino aliquae transirent moverenturque...caelo in hereditate cunctis
relicto, si quisquam, qui cretionem eam caperet, inventus esset.''}
(Plinius, Naturalis Historia II.24).

This work has been supported through the ASI grant 59/1997.}

\appendix
\section{Metal content and main sequence position}

We wish to compare theoretical and observed $\Delta (B-V)/\Delta [Fe/H]$.
As reference loci we take isochrones published recently by our
group. For Population II we adopt the results by D'Antona et al. (1997),
especially computed for dating globular clusters; for Population I we use
either the main sequence location for $Z=0.01$\ and 0.02 from D'Antona \&
Mazzitelli (1997), or the homogeneous grid of results by Ventura et al.
(1998). In all the models the helium abundance is scaled according to the law
$\Delta Y/\Delta log Z$ = 3. The theoretical tracks are transformed to the
observational plane $M_v$, \BMV\ through Kurucz (1993) relations, except
isochrones by Ventura et al. (1998), transformed through Castelli (1998)
relations.

In Fig. A1 main sequence objects from the metallicity sample are shown, for
various [Fe/H] intervals. We estimated by eye the colours at \Mv\ = 6; the
estimated error is about $\pm$ 0.03 mag.  Around \Feh\ = 0, the
ratio $\Delta (B-V)/\Delta [Fe/H]$ is about 0.20, decreasing to 0.16 for
lower metallicities. Towards very low \Feh\ ($\leq$ -1.2), the ratio seems to
decrease further, but the objects are too few for a meaningful estimate.

The tracks by D'Antona et al. (1997 and extensions to higher \Z) give, at
\Mv\ = 6 and for very low \Z\ ($10^{-4} -- 10^{-3}$), $\Delta (B -
V)/\Delta [Fe/H]$ = 0.08; for higher \Z, up to 0.01, the ratio is about
0.16. Beyond the solar metallicitiy, the ratio increases to about 0.25.
The theoretical relation is also shown in Fig. A1. We see that
changes in metal content are mirrored by shifts in the main sequence average
colours, in reasonable agreement with theoretical predictions. We conclude
that, with some caution, it is possible to follow changes in populations
(with differing metal contents) through changes in CM diagram main sequence
loci.

\begin{figure}
\hspace{0.8cm}
\caption{The boxes represent the color of the main sequence at \Mv=6
for different metallicity bins. The theoretically expected values from
D'Antona et al. (1997) complemented by the values of D'Antona and Mazzitelli
(1997) are shown by dots.} \label{figA1}
\end{figure}

Another check on the behaviour with metallicity of Hipparcos objects has
been performed considering the subdwarfs in the catalogue studied
by Gratton et al. (1997, Table 1). These objects have Hipparcos parallaxes
such that $\Delta\pi / \pi < 0.12$, and abundances derived from analysis
of high dispersion spectra. The location of these stars is shown in Fig. A2,
where we adopt as reference lines isochrones of 10 Gyr for metal poor stars,
and of $10^8$ yr for the metal rich population. On the whole, the agreement
between theory and observations turns out satisfactory, but with some
exceptions which caution to apply average conclusions to isolated cases,
without a specific analysis.

\begin{figure}
\hspace{1cm}
\caption{The subdwarfs from Table 1 in Gratton et al. 1997 are shown
together with four $10^{10}$ yr isochrones (\Z\ = 1 and 3 $10^{-4}$, 1 and 4
$10^{-3}$ from left to right) and two $10^8$ yr isochrones (\Z\ = 0.01 and
0.02).}
\label{figA2}
\end{figure}

One of the difficulties is given by the not negligible dependence on the
adopted transformation of the location in CM diagrams of theoretical tracks.
In any case, we are interested in relative positions and the following
considerations should remain substantially valid also after a shift
of the zero point on the observational plane.

Within the errors, 18 out of the 20 more metal poor stars ([Fe/H] $\leq$ -1)
lie on or at the left of the \Z\ = $10^{-3}$ isocrone (\Feh\ = -1.26),
therefore in good overall agreement with theoretical expectations. Besides,
if we consider only the subdwarfs which, within errors, fall on the main
sequence of $10^{-3}$ (14 objects) and compute the average of their \Feh\
values, we find -1.28, that is, almost exactly \Z\ = $10^{-3}$.

At the same time, the four objects lying on the very metal poor tracks (1 and
3 $10^{-4}$) give an average of \Feh\ = -1.69, that is, \Z\ = 3.7 $10^{-4}$,
slightly larger than the comparison tracks. On the other hand, some intrinsic
uncertainty appears still present even in the high quality data from Gratton
et al.: for example, of the two most metal poor objects (\Feh $= -1.91$\ and
$-1.92$), one falls between the 1 and 3 $10^{-4}$ tracks and the other on the
$10^{-3}$ track (all these metal poor tracks are, in any case, very close).

Similar difficulties are found when considering stars with \Feh $>-1.0$.
The average of the three objects with $-1<$ \Feh $<-0.7$\ gives -0.88
(\Z\ = 2.3 $10^{-3}$), while they lie between the tracks of \Z\ = 4
$10^{-3}$ and $10^{-2}$. As for the 11 most metal rich subdwarfs in Fig.
A2 ($-0.7<$ \Feh $<-0.4$, average \Z\ = 5 $10^{-3}$), they are found
(within errors) between the tracks for \Z\ = 4 $10^{-3}$ and 0.02.

Some of these discrepancies may be reduced by taking into account
$\alpha$--enhancement (Salaris et al. 1993), which increases \Z\ estimates
beyond those given by the iron peak elements. Still, the dispersion in
colour observed for stars with similar \Z\ suggests that the evaluation of
metal content suffers from larger errors than currently indicated.

In any case, let us consider more in details two rather well defined groups,
the one with $-0.7 <$ \Feh $<-0.4$\ and the one with $-1.51 \leq $ \Feh
$\leq -1.0$. We evaluated the average value of their \BMV\ at \Mv\ = 6 mag,
by shifting the star position according to $\Delta (B-V)$/$\Delta M_{\rm
v}$. This quantity was obtained from theoretical sequences in the interval
5.5 $<$ \Mv\ $<$ 6.5 mag, where evolutionary effects are negligible, and
turned out to be $\sim$ 0.19.

For the first group we found an average \Feh $=-0.53$ and an average \BMV\
= 0.791; for the second one, -1.28 and 0.649, respectively. Therefore,
$\Delta (B-V)$/$\Delta [Fe/H]$ = 0.19. In the metallicity interval $10^{-3}$
$\leq$ $Z$ $\leq$ $10^{-2}$, theoretical isochrones (D'Antona et al. 1997)
transformed with Kurucz (1993) give for this ratio 0.16. The larger
observational separation in metallicity is due to the fact that the more
metal rich group has a substantially redder colour than expected from its
\Feh, while the more metal poor one is only slightly redder than
predictions.

Again, changes in metal content give rise to changes in main sequence
colours in overall agreement with model predictions. This supports the
conclusions reached before, on the basis of the whole metallicity sample,
that it is justified to attribute variations in main sequence loci to
variations in metal content.

\begin{thebibliography}{}

\bibitem[]{} Badiali M., Cardini D., Emanuele A., et al.,
1997, in ESA Symposium Hipparcos Venice '97, ESA SP-402, 661
\bibitem[]{} Barbier-Brossat M., Petit M., Figon P., 1994, A\&AS 108, 603
\bibitem[]{} Bartkevicius A., Bartasiute S., Lazauskaite R., 1997, in ESA
Symposium Hipparcos Venice '97, ESA SP-402, 343
\bibitem[]{} Binney J.J., Dehnen W., Houk N., Murray C.A., Penston M.J.,
1997, in ESA Symposium Hipparcos Venice '97, ESA SP-402, 473
\bibitem[]{} Carney B.W., Latham D.W., Laird J.B., 1990, AJ 99, 572
\bibitem[]{} Carney B.W., Latham D.W., Laird J.B., Aguilar L.A., 1994, AJ
107, 2240
\bibitem[]{} Castelli F., 1998, in Views on Distance Indicators, eds. M.
Arnaboldi, F. Caputo, A. Rifatto, Mem. SAIt. 69, 165; tables available at
http://cfau5.harvard.edu
\bibitem[]{} Cayrel R., Lebreton Y., Perrin M.N., Turon C., 1997, in ESA
Symposium Hipparcos Venice '97, ESA SP-402, 219
\bibitem[]{} Cayrel de Strobel G., Soubiran C., Friel E.D., Ralite N.,
Francois P., 1997, A\&AS 124, 299
\bibitem[]{} Chiba M., Yoshii Y., 1998, AJ 115, 168
\bibitem[]{} D'Antona F., Mazzitelli I., 1997, in Cool Stars in Clusters and
Associations, eds. R.Pallavicini, G.Micela, Mem.S.A.It. 68, 807
\bibitem[]{} D'Antona F., Caloi V., Mazzitelli I., 1997, ApJ 477, 519
\bibitem[]{} Duflot M., Fehrenbach C., Mannone R., Burnage R., Genty V.,
1995a, A\&AS 110, 177
\bibitem[]{} Duflot M., Figon P., Meyssonnier N., 1995b, A\&AS 114, 269
\bibitem[]{} Eggen O.J., Lynden-Bell D., Sandage A.R., 1962, ApJ 136, 748
\bibitem[]{} ESA 1997, The Hipparcos and Tycho Catalogues, ESA SP-1200
\bibitem[]{} Fehrenbach C., Duflot M., Genty V., Amieux G., 1996, Bull. Inf.
CDS 48, 11
\bibitem[]{} Fehrenbach C., Duflot M., Mannone C., Burnage R., Genty V.,
1997, A\&AS 124, 255
\bibitem[]{} Gilmore G., Reid I.N., 1983, MNRAS 202, 1025
\bibitem[]{} Gilmore G., Wyse R.F.G., 1985, AJ 90, 2015
\bibitem[]{} Gliese W., 1969, Catalogue of Nearby Stars, Veroff. Astron.
Recken-Inst. Heidelberg No. 22
\bibitem[]{} Gliese W., 1974, A\&A 34, 147
\bibitem[]{} Gliese W., Jahreiss H., 1979, A\&AS 38, 423
\bibitem[]{} Gratton R., Fusi Pecci F., Carretta E., et al.,
1997, ApJ 491, 749
\bibitem[]{} Houk N., 1978, Catalogue of Two-Dimensional
Spectral Types for the HD Stars, Vol. 2, Dept. of Astronomy, Univ. of
Michigan, Ann Arbor
\bibitem[]{} Houk N., 1982, Catalogue of Two-Dimensional
Spectral Types for the HD Stars, Vol. 3, Dept. of Astronomy, Univ. of
Michigan, Ann Arbor
\bibitem[]{} Houk N., Cowley A.P., 1975, Catalogue of Two-Dimensional
Spectral Types for the HD Stars, Vol. 1, Dept. of Astronomy, Univ. of
Michigan, Ann Arbor
\bibitem[]{} Houk N., Smith-Moore M., 1988, Catalogue of Two-Dimensional
Spectral Types for the HD Stars, Vol. 4, Dept. of Astronomy, Univ. of
Michigan, Ann Arbor
\bibitem[]{} Ibata R.A., Gilmore G., Irwin M.J., 1994, Nature 370, 194
\bibitem[]{} Kerr F.J., Lynden-Bell D., 1986, MNRAS 221, 1023
\bibitem[]{} Kurucz R.L., 1993, CD-ROM 13 and CD-ROM 18
\bibitem[]{} Lebreton Y., Perrin M.N., Fernandes J., et al.
1997, in ESA Symposium Hipparcos Venice '97, ESA
SP-402, 379
\bibitem[]{} Luyten W.J., 1976, LHS Catalogue, Minneapolis, Univ. of
Minnesota (and revised version).
\bibitem[]{} Luyten W.J., 1979, NLTT Catalogue, Minneapolis, Univ. of
Minnesota
\bibitem[]{} Majewski S.R., 1993, ARA\&A 31, 575
\bibitem[]{} Norris J., 1986, ApJS 61, 667
\bibitem[]{} Norris J., Bessel M.S., Pickles A.J., 1985, ApJS 58, 463
\bibitem[]{} Perryman M.A.C., Turon C., Arenou F., et al., 1989, ESA
SP-1111, Vol. II, p. 89
\bibitem[]{} Salaris M., Chieffi A., Straniero O., 1993, ApJ 414, 580
\bibitem[]{} Sandage A., Fouts G., 1987, AJ 92, 74
\bibitem[]{} Str\"omberg G., 1925, ApJ 61, 363
\bibitem[]{} Turon C., Creze M., Egret D., et al., 1992, ESA SP-1136
\bibitem[]{} Ventura P., Zeppieri A., Mazzitelli I., D'Antona F., 1998,
A\&A 334, 953
\bibitem[]{} Wyse R.F.G., Gilmore G., 1986, AJ 91, 855
\bibitem[]{} Yoshii Y., Saio H., 1979, PASJ 31, 339
\end{thebibliography}
\end{document}